# CONSCIOUSNESS AND THE PROBLEM OF QUANTUM MEASURMENT

Chris Allen Broka
(chris.broka@gmail.com)

**Abstract**

A variant of the von Neumann-Wigner Interpretation is proposed. It does not make use of the familiar language of wave functions and observers. Instead it pictures the state of the physical world as a vector in a Fock space and, therefore not, literally, a function of any spacetime coordinates. And, rather than segregating consciousness into individual points of view (each carrying with it a sense of its proper time), this model proposes only unitary states of consciousness, $Q(t)$, where $t$ represents a fiducial time with respect to which both the state of the physical world and the state of consciousness evolve. States in our world's Fock space are classified as either 'admissible' (meaning they correspond to definite states of consciousness) or 'inadmissible' (meaning they do not). The evolution of the state vector of the world is such as to always keep it restricted to 'admissible' states. Consciousness is treated rather like what Chalmers has called an "m-property." We try to show that problems with the quantum Zeno effect do not arise from this model. The measurement of degenerate states is also discussed, as is the EPR paradox. Finally, it is suggested that the universe is far more likely to contain conscious observers than not.

## Introduction.

The relationship between consciousness and quantum mechanics has long been discussed in a serious context. Schrödinger (1), Wigner (2), and von Neumann (3) were, early on, most associated with these lines of investigation. And Everett's Relative State Interpretation is very much about our states of consciousness. Still, it seems that little effort has gone into developing mathematical formalisms appropriate to the description of this relationship. Perhaps this is because mental states seem to be mathematically indefinable. The von Neumann-Wigner idea has never been very popular with physicists, who mostly hold to somewhat Instrumentalist views; the wave function collapses because macroscopic, complicated, things like detectors cannot exist in superposed states. How big macroscopic things are, or how complex something must be to be complicated, are questions left largely to the imagination. We do like the von Neumann-Wigner Interpretation for two reasons: For one thing, it is intuitively obvious to us why our sensations cannot exist in superposition with one another. The same cannot be said for purely physical states that happen to be large or complex. Also, it provides a natural role for consciousness in physics; we would not like to regard consciousness as a mere epiphenomenal innocent-bystander (although some readers probably would).

Recognizing that conscious states never seem to exist in superposition, Chalmers and McQueen (4, 5) introduced the very clever idea of m-properties. These properties cannot exist in superposition. They take consciousness to be such a property and assign it an operator (which we will designate m) that measures it and projects states – they treat consciousness very much like an ordinary physical observable. The present work tries to extend this idea and resolve some of the problems it engenders. Attention is paid to the quantum Zeno effect.

## States of the World.

Designate the state of the physical world as $|\Psi(t)\rangle$. Such states are to be understood as vectors in a Fock space. Basis vectors in this space are constructed from the vacuum state $|0\rangle$ by the repeated application of creation operators appropriate to the various kinds of particles that inhabit our universe. The parameter $t$ recognizes that this state evolves with respect to a fiducial time that we can identify with a particular Lorentz frame. It is essential to recognize that $|\Psi(t)>$ is *not* a function of the spatial coordinates (x, y, z). We will work in the Dirac Interaction Picture. Here we regard our Fock space as built using the creation operators appropriate to free, non-interacting particles and write $i\ \partial_t\ |\Psi(t)> \ = \ H'\ |\Psi(t)>$ where $H'$ is the Hamiltonian describing the interactions amongst these particles. The complete Hamiltonian for the system is written $H = H_0 + H'$. $|\Psi(t)>$, in fact, describes the entire world without making arbitrary distinctions between observers, detectors, and things detected.

## States of Consciousness.

Suppose that, at any time $t$, the conscious state of the universe can be designated as $Q(t)$. $Q(t)$ describes the *qualia* — the totality of sensations experienced by any and all consciousness anywhere at that time; that could include physicists recording a quantum measurement or worms tasting sugar in a pond on a distant planet.

There should be no segregation of consciousness into any set of individual observers. We will just designate it $Q(t)$. This will prevent us from trying to describe the world in terms of separable and independent "wave functions," one for your brain, my brain, or other things; it would be difficult having to make sense of trillions of independent consciousnesses. While it is true that we seem to experience our individual conscious states independently, we have to argue that this apparent separateness is merely an illusion born of facts like "I" can remember my memories but not "yours." And "I" can experience my sensation of blue but not "yours." These facts are indisputable, and surely interesting. But they only obscure matters if we wish to study the problem at hand. And we do not want to be misinterpreted as proposing anything mystical here. The important point is mathematical — we will regard consciousness (in its totality) as something that can be indexed by a *single* parameter $t$.

There will be a set of Fock space states, $|C_i>$, that correspond to what we will call *definite states of consciousness*. By this we mean that the associated $Q(t)$ specifies a unique and unambiguous state of awareness possessed by the totality of sentient observers. Any state that does not correspond to a definite state of consciousness will be called a mixed state. Definite states of consciousness are *admissible* states in this theory. Mixed states are *inadmissible* and cannot be allowed to occur. (By writing things this way it might appear that we regard the $Q_i$ and $|C_i>$ as constituting discreet, denumerably infinite sets. This is, of course, not the case. Both are, properly, to be regarded as continua, as is the set of inadmissible states. This notation is just simpler to work with.)

Regarding the nature of the $|C_i>$s one thing is obvious; they are highly degenerate with respect to the qualia – states. It would, for instance, make no difference to the overall state of consciousness whether an electron had been created in a region with no observers. And we must recognize a sort of null state of consciousness — a state where there just aren't any sentient observers at all. (The state vector a few seconds after the big bang would correspond to such a state. So would many others.) We must picture our Fock space as broken up into many separate subspaces, some with a particular $Q_i$ that designates the unique conscious experience corresponding to it, and others corresponding to no definite experience at all.

The time variable that appears in $|\Psi(t)>$ and $Q(t)$ requires a comment. As it pertains to the former case it causes no problems with relativity since the equations that determine the evolution of $|\Psi(t)>$ are, themselves, relativistically invariant — $t$ only represents an arbitrary choice of Lorentz frame. $Q(t)$ might cause a problem, however. By choosing not to regard consciousness as broken up into separate observers (each of which needing to be assigned its own proper time) we have more-or-less forced ourselves to select a particular set of space-like hypersurfaces to designate the various '*t*'s. Now perhaps because consciousness is a non-material sort of thing such a violation of relativity is permissible — we can't be sure how physics treats non-material things. But it is essential that we construe *Q(t)* in such a way as to end up with no *physical* violations of relativity.

## The Evolution of these States with Time.

Taking no account of consciousness we could picture $|\Psi(t)>$ evolving according to $i\, \partial_t \,|\Psi(t)> \;=\; H'(t)\,|\Psi(t)>$ where $H'$ designates the interaction operator for our world. ($H'(t) = \int \mathcal{H}'(x, t)\, d^3\boldsymbol{x}$ where $\mathcal{H}'(x, \;)$ is the corresponding Hamiltonian density operator). We assume normal ordering. All operators and state vectors are being represented in the Dirac Interaction Picture. There would, in consequence, exist a unitary operator, $\mathrm{U}(t_2, t_1)$, having the property that $|\Psi(t_2)> = \mathrm{U}(t_2, t_1)\, |\Psi(t_1)>$ . Let us imagine the world at time $t$ being in a definite state of consciousness. Now the $|C_i>$s are in no necessary way eigenstates of the Hamiltonian so things could quickly evolve into a situation where $|\Psi(t)>$ became inadmissible. But this is never what we seem to experience; our common awareness appears unconfused and composed of a well-defined succession of qualia. Reality will only tolerate definite states of consciousness — that is to say $|\Psi(t)>$ must *always* be admissible.

We can arrange for this to happen by amending the previous equation for the time-evolution of $|\Psi(t)>$ to also require $\mathfrak{S}\,|\Psi(t)> \;=\; |\Psi(t)>$ where $\mathfrak{S}$ is a (non-linear) operator having some interesting properties:

1) If $|\Psi(t)>$ is admissible it does nothing. The state is completely unaffected.

2) If $|\Psi(t)>$ is not admissible it will look at all the amplitudes $<C_i|\Psi(t)>$ for every existing $<C_i|$. It will square these amplitudes and, using these values as *relative* probabilities, convert $|\Psi(t)>$ into one of the $|C_i>$ at random.

$\mathfrak{S}$ functions as a projection operator taking mixed states into definite states of consciousness. We give up the idea of a unitary time-evolution operator. Such an operator has an inverse. We cannot go backwards in time according to $\mathfrak{S}$ since the decision how to go forward is made at random. This imparts a natural directionality to time. $\mathfrak{S}^2 = \mathfrak{S}$ and $\mathfrak{S}$ has no explicit time dependence. The qualia-state is assumed independent of phase so, if $|\Psi(t)>$ corresponds to a particular $Q_i$, $e^{i\,\theta}\, |\Psi(t)>$ will correspond to it also. We also recognize a null-state of consciousness corresponding to the existence of no qualia whatsoever. Many $|\Psi>$s will correspond to it. These are considered admissible.

It is worthwhile to consider the difference between the present idea and the m-property theory. Chalmers and McQueen appear to regard the m-operator (which measures consciousness) as behaving like any standard measuring operator. It seems to play an active role in projecting the state into one of its eigenstates. But for us, $\mathfrak{S}$ plays not so much an active as a *permissive* role — it distinguishes admissible state vectors from inadmissible ones. Here is a picturesque metaphor: We are accustomed to thinking of the Fock space in which our reality lives as something like an infinitely extended, infinite-dimensional block of Cheddar cheese. We, instead, picture it more like a block of Swiss cheese — it is full of holes. The cheese contains the state vectors that represent definite states of consciousness. The holes contain the mixed states. Ordinarily $|\Psi(t)>$ evolves, under the action of U, so as to remain inside the cheese. $\mathfrak{S}$ does nothing at all. But sometimes (perhaps due to the intervention of experimental physicists) it tries to move into one of the holes. At the instant it does this $\mathfrak{S}$ corrects the situation by projecting it back into the cheese. But $\mathfrak{S}$ is a very "lazy" operator. It tries to project the errant state into the most similar states available in the cheese. This idea would run into serious trouble if the state vector were to evolve into a $|\Psi(t)>$ that existed in the "cheese" but was about to enter a "hole" if $|\Psi(t)>$ was orthogonal to *every other* $|C_i>$. $\mathfrak{S}$ would have no choice but to project the state back into itself and reality would hang up there forever. Since the universe seems to have been evolving successfully for about 14 billion years, we assume that our "Swiss cheese" is so densely packed with $|C_i>$s that this situation never arises. $\mathfrak{S}$ might have to project the state many times before it got lucky and found a way around the hole. But we assume $\mathfrak{S}$ functions instantaneously. We would, ourselves, never notice a thing.

It might be objected that our "fiducial time" violates relativity by introducing a preferred Lorentz frame. It does do this, of course, but only in relation to qualia. We are Property Dualists and do not think that qualia are, in any sense, physical things. We believe that they can violate relativity as much as they want provided that no *physically observable* contradictions of relativity result. Since we have assigned consciousness only a permissive (as opposed to active) role, we do not see a problem here. I think it would be very difficult to use this idea to construct any physical experiment that would show a violation of relativity. (But it is hard to prove a negative.)

## The Anatomy of a Measurement.

Consider a very simple experiment in which an electron is sent through a Stern-Gerlach apparatus. It can be prepared as either spin-up or spin-down or in any superposition of these states. If it comes in spin-up it always veers up and strikes a detector that causes a light, originally blue, to shine green. If it is down it goes the other way and a red light is triggered. This device, the electron whose spin it measures, and an observer, constitute a physical universe described by $|\Psi(t)>$. The conscious states of this universe, we will imagine, belong to this single observer whose only possible states of awareness are 1) seeing a green color, 2) seeing a red color, or 3) seeing a blue color. So the space in which the conscious state of the universe is a vector contains three sub-spaces – one corresponding to each of the above possibilities. For simplicity, we assume that the blue, red, and green states are mutually orthogonal. Since this world is simple, we think that we can get away with describing it in a simple manner. Let us describe its initial state as $|+, B>$ where + says that our electron is spin-up. 'B' simply says that the rest of the measurement system (observer and all) are in their initial state. When the spin-up electron is detected the initial state evolves into $|+, G>$ meaning that this new state is an admissible one corresponding to the qualia 'seeing a green color.' (If the electron had been spin-down we would have ended up with a red qualia and a state $|-, R>$.) If things happen to start out as $(|+, B> + |-, B>)/\sqrt{2}$ our system will, obviously, evolve into a superposition of states which is no longer admissible. It would correspond to the observer seeing, simultaneously, a green and a red qualia. Wigner described this situation as "absurd." Since, according to the above-mentioned principle, reality cannot tolerate any inadmissible state it is necessary that $\mathfrak{S}$ project it into either $|+, G>$ or $|-, R>$ with 50% probability. Let us make it clear that no wave function collapses. Instead, a state vector $|\Psi(t)>$ – which is *not* a function of the spatial coordinates (x, y, z) – tries to evolve into an inadmissible state (in Fock space). $\mathfrak{S}$ immediately corrects this by projecting $|\Psi(t)>$ back into only one of the two *definite* states of consciousness available to it.

It could be objected that our projection scenario would not, in general, give us the Born Rule. This is entirely correct. The various $|C_i>$s into which $|\Psi(t)>$ might find itself projected will not usually be orthogonal to one another — in reality, $|\Psi(t)>$ describes not only the Stern-Gerlach apparatus and its electron but, also, the enormous, possibly infinite, number of other particles that exist in our universe. Our small experiment represents only a miniscule "part" of $|\Psi(t)>$ and we would expect $| < C_i | \Psi(t) > |^2$ to be nearly 1 irrespective of our experiment provided that projection into $|C_i>$ did not involve much change to the rest of the universe. But this does not reflect the way in which actual quantum measurements are performed.

Suppose some experimentalists wanted to determine the energy of a superposed photon state consisting of a red photon and a green one by making a von Neumann measurement. They would evacuate their apparatus, surround it with thick lead shielding, and, perhaps, place it in an abandoned salt mine. They would do everything possible to insulate it from any interaction with the rest of the universe. They try to create a sort-of "mini-universe" consisting of only their photon and detector. To the extent they succeed, their red-photon and green-photon states really can be considered as orthogonal and the Born Rule will be recovered. If they tried to perform their experiment outside on a sunny day they would, of course, not find the Born Rule to work at all. They would probably just be observing random photons coming from the Sun.

For the Born rule to obtain the state vectors describing the final (admissible) states must be orthogonal or, at least, very nearly so. In the case of the isolated experiment we can imagine the detector system being composed of a vast number of particles, none of which are photons. Adding a single red photon will change all the basis vectors into new ones. Adding a single green photon would change them into a completely different set of new basis vectors and the two states truly would be orthogonal. Had the detector state already been contaminated with an enormous and indefinite number of green and red photons, adding one extra red one would be very unlikely to produce a state orthogonal to the one produced by adding one extra green one. A similar analysis could be applied to the Stern-Gerlach experiment discussed above.

We would want to assume that there is some algorithm or "test" that can be applied to a state vector in Fock space to determine whether it represents a definite (admissible) state of consciousness or a mixed (inadmissible) one, and, if an admissible one, to what definite qualia-state it corresponds. The details are hard to pin down, but we can say a few things. The superposition of two or more *non-null* admissible states which do not all correspond to the same qualia-state is to be considered inadmissible (e.g. the "absurd" green-red qualia). If they do all correspond to the same qualia-state, the superposition will be admissible and correspond to the one common qualia. The superposition of an admissible state and an inadmissible one might be admissible or not. Consider (|G> - |R>)/$\sqrt{2}$ (inadmissible) and |R> (admissible). A superposition of one times the first and $1/\sqrt{2}$ times the second would be |G>, which is admissible. Other superpositions would not be. The same holds for the superposition of two (or more) inadmissible states. Null states represent an interesting problem. We can easily see this if we notice that the (complete) basis states of our Fock space, which each describe the existence of an integral number of particle species having definite momenta, must all be admissible null states of consciousness. Having only particles with definite momenta (therefore being "everywhere" in space), they cannot be otherwise since, by their nature, they cannot individually describe localized material things like planets or animals. But the admissible state vector that is our own, conscious, reality can be written as the superposition of an enormous number of these basis vectors. So can any inadmissible state vector. A superposition of a null state and an admissible (or inadmissible) one might be admissible or not.

We must comment on the notion of *continuity* in our theory. As applied to $|\Psi(t)>$ its validity is obvious and, so long as no quantum measurements are made, $|\Psi(t)>$ evolves continuously. The same cannot be said for $Q(t)$, however. $\frac{Q(t+\epsilon)-Q(t)}{\epsilon}$ makes no mathematical sense at all. Even if $|\Psi(t)>$ evolves continuously we suppose that the corresponding qualia-state might, at some point, change discontinuously into an altogether different one. Two state vectors, even if differing only infinitesimally, might correspond to very different qualia.

## The Quantum Zeno Effect.

Since $\mathfrak{S}$ measures the state vector constantly a concern may arise regarding the quantum Zeno effect (6); if the state is always being observed can it really ever go from blue to green or red? This problem has been discussed carefully by Chalmers and McQueen in their consideration of m-properties.

Consider a spin-up electron moving through the Stern-Gerlach apparatus but still far from the detector. $|\Psi(t)\rangle$ will, of course, be changing. But it will always remain one of the many (admissible) B qualia-states and $\mathfrak{S}$ will affect it in no way. As the electron moves along toward the detector, a very frequent measurement of its position might, indeed, stop it somewhat from moving. But a very frequent measurement of the blueness of the light will not affect its motion whatsoever.

At $t_d$ it encounters the measuring device. $|\Psi(t_d)\rangle$ is now an admissible state corresponding to a green qualia. Throughout this entire process $\mathfrak{S}$, for all its observing, *has done not a single thing to the state vector*. Now, if the electron had been in a superposed state when it hit the detector, $\mathfrak{S}$, at $t_d$, would have immediately projected things into either of the two possible outcomes. But, still there is no Zeno effect to be noticed in the process. Why? $\mathfrak{S}$ measures things all the time but, as things proceed along here, $|\Psi(t)\rangle$ *always* remains an admissible state.

Say that $|\Psi(t_d)\rangle$ is just about to enter a "hole." $\langle\Psi(t_d)|\Psi(t_d)\rangle = 1$ at this time so it might seem as if $\mathfrak{S}$ would have to project it back into itself. Not at all. Suppose there was another admissible state, $|\Phi\rangle$, that was such that $|\langle\Phi|\Psi(t_d)\rangle|^2 = .999$. This is altogether possible and $|\Phi\rangle$ might even correspond to an entirely different qualia-state than $|\Psi(t_d)\rangle$. There would now be an almost equal probability of projection into either state. (There might even be 1000 such $\Phi$-type states!) For us, the various squared amplitudes must be interpreted as *relative*, not absolute, probabilities. Another important point to bear in mind is the following: For the Quantum Zeno effect to obtain it is assumed that the system evolves unitarily for some short time, gets measured, evolves unitarily for a little more time, gets measured again, and so on. We propose nothing of the sort. In this model $\mathfrak{S}$ acts only once and it does so instantaneously.

A bigger problem for our theory is why the quantum Zeno effect can be demonstrated at all; it is, indisputably, a real thing. Excited beryllium ions have been prevented from decaying by pulsing them frequently with light to detect if they are still in their original state (7). This is a perfectly valid experiment since measuring the ion as excited completely precludes any possibility of its being decayed. But the rate of testing of the ion (as fast as 250 times/sec) greatly exceeds what any human brain could consciously process. It appears that human consciousness cannot be affecting the ion. Is the detector supposed to be conscious too?

We need to consider this process more carefully. Suppose the detector makes only one measurement at time T. T is very small relative to the rate of decay. Initially the ion is in its undecayed state (U) and the detector is still in its initial state (I). We will write its state vector as $|U, I\rangle$. At time T the measurement is made. Consciousness notices nothing and nothing projects. But the detector records its activities on a strip of magnetic tape. The state is now $\alpha\,|U, I, I\rangle + \beta\,|D, I, A\rangle$, where A means the detector has registered a decay, $\alpha \approx \sqrt{1 - \frac{T^2}{\tau_z^2}}$ and $\beta \approx \frac{T}{\tau_z}$ (up to irrelevant phases). $\frac{1}{\tau_z^2} = (|\langle U|H|U\rangle|)^2 - \langle U|H^2|U\rangle$. It is assumed that the possible states are complete and orthonormal basis vectors spanning the Hilbert space of our system. If consciousness now looks at it, the probability of finding the ion undecayed will be $\alpha^2$. Suppose, instead, that two measurements are made, one at T/2 and another at T. The state now becomes $\alpha^2\,|U, I, I, I\rangle + \alpha\,\beta\,|D, I, I, A\rangle + \beta\,|D, I, A, A\rangle$ where $\alpha \approx \sqrt{1 - \frac{T^2}{4\,\tau_z{}^2}}$ and $\beta \approx \frac{T}{4\,\tau_z}$ (measurement does nothing to an already decayed state and the state remains normalized). Everything, the tape included, is still in superposition. Now consciousness looks at it. It wants to know the probability of finding |U, I, I, I>, i.e. that the particle has not decayed. This is $\left(1 - \frac{T^2}{4\,\tau_z{}^2}\right)^2$ which is larger than the previous result. This is what it would have been had the sequential measurements each actually projected (collapsed) the state. It does not matter whether they do or whether we simply look at (and collapse) the strip of tape after the experiment has concluded. We could further slow the decay by dividing T into a larger number of measurement intervals. The decay is inhibited just as Zeno would have it. Although they might seem to do nothing, the measurements have altered the trajectory of the state vector through our Fock space. Had they not been performed, the system would have evolved differently.

## Causality.

We must attribute one more property to $\mathfrak{S}$. $|\Psi(t)>$ is in no way a function of the spatial coordinates. But it does contain all the information we can have regarding what is going on where. (For example, a Newton-Wigner state, $\sum_k e^{i\,k\cdot x} a_k^\dagger |0>$ (suitably normalized), describes a universe with one particle localized at $\boldsymbol{x}$.) To take account of relativity we must make an additional stipulation:

3) When $\mathfrak{S}$ projects an inadmissible state into an admissible one that new state must be such that it describes the same physics as the old one *outside the future light cone of the measurement event*.

Otherwise, physical information regarding the outcome of the measurement could travel faster than light. We cannot allow for this.

## Degenerate States and the EPR Paradox.

Provided our apparatus is effectively shielded from all outside influences, and supposing that our measurement does not destroy the system, we will usually end up with a normal von Neumann measurement and the Born rule. An interesting problem arises if our system is degenerate with respect to eigenvalues that could be obtained through our measurements. What does the state collapse into then? Dirac (8), von Neumann (9), and Lüders (10) have tried to give us answers. We note that our above-described protocol gives predictions that differ considerably from those of the above-mentioned authors. For recent discussions see (11, 12, 13, 14).

Suppose we have a system that can be characterized by two observables. Here we will use energy and spin. Suppose it exists in a Hilbert space spanned by three orthonormal basis states |I>, |II>, and |III> where:

4) |I> = |Energy = $E_1$, spin = ->, |II> = |Energy = $E_2$, spin = ->, |III> = |Energy = $E_2$, spin = +>.

Let there be a pure state we will describe as:

5) $|\Psi_0> = c_1\,|I> + c_2\,|II> + c_3\,|III>$ with $|\,c_1\,|^2 + |\,c_2\,|^2 + |\,c_3\,|^2 = 1$.

Suppose we measure its energy. (We must find a definite value; a pointer cannot be perceived to point at two energy values simultaneously.) There is a probability, $|\,c_1\,|^2$, that we will find $E_1$. If we do, the state will be projected into |I>. If we, subsequently, measure the spin we are certain to find it -. (We assume that our states are eigenstates of the Hamiltonian and will only change by meaningless phases as time goes on.) On this point we and the other interpretations all agree. Suppose we find $E_2$. Then into what state is it projected? For Lüders the answer is simple: It is projected into $(c_2\, |\mathrm{II}\rangle + c_3\, |\mathrm{III}\rangle)/\sqrt{|c_2|^2 + |c_3|^2}$. We can then measure the spin. According to Lüders we have a probability $|c_1|^2$ of ending up in |I>, $|c_2|^2$ of ending up in |II>, and $|c_3|^2$ of ending up in |III>. There is nothing obviously unreasonable about this result. But our protocol leads to very different conclusions.

Suppose we measure $E_2$. *All* linear combinations of |II> and |III> are $E_2$ energy eigenstates and, therefore, admissible in terms of an energy measurement. The state could project into any of them in the manner suggested above. Let us write the most general combination as:

6) $\psi_a(\beta, \theta) = e^{i\theta}\sqrt{1 - \beta^2}\; |\mathrm{II}\rangle + \beta\, |\mathrm{III}\rangle$ where $\beta$ is real and between 0 and 1. $0 \le \theta \le 2\pi$.

The above retains an ambiguity since it can be multiplied by an arbitrary phase and nothing is changed. We do not have to worry about this as the same ambiguity applies to |I>. Since all we are interested in are the *relative* probabilities the effect of taking this into account nets out to zero. The total probability of measuring $E_2$ is given by:

7) $\int_0^1 \int_0^{2\pi} \left| \langle \Psi_0 \middle| \psi_a(\beta, \theta) \rangle \right|^2 \mathcal{M}(\beta, \theta)\, d\theta\, d\beta$ where $\mathcal{M}(\beta, \theta)$ represents a *measure* over the $(\beta, \theta)$ "space."

This gives us:

8) Total probability of $E_2 = \int_0^1 \int_0^{2\pi} \left| c_2^*\, e^{i\theta} \sqrt{1 - \beta^2} + c_3^*\, \beta \right|^2 \mathcal{M}(\beta, \theta)\, d\theta\, d\beta$.

If $\mathcal{M}(\beta, \theta)$ is assumed to be independent of $\theta$ the $e^{i\theta}$ containing terms integrate out and we are left with:

9) Total probability of $E_2 = 2\pi \int_0^1 \left( |c_2|^2 \left(1 - \beta^2\right) + |c_3|^2\, \beta^2 \right) \mathcal{M}(\beta)\, d\beta$.

But what to choose for $\mathcal{M}(\beta)$? Our measure represents the number of admissible states in the interval $d\beta\, d\theta$. Our $(\beta, \theta)$ "space" is, in polar coordinates, just the unit disk. The volume element here is $\beta\, d\beta\, d\theta$. This suggests we should try $\mathcal{M}(\beta, \theta) = \beta\, \mathcal{M}_0$.

10) Total probability of $E_2 = \int_0^1 \int_0^{2\pi} \left| c_2^*\, e^{i\theta} \sqrt{1 - \beta^2} + c_3^*\, \beta \right|^2 \beta\, \mathcal{M}_0\, d\theta\, d\beta$.

If we set $\mathcal{M}_0 = 2/\pi$ we find that the total probability of ending up with $E_2$ is $|c_2|^2 + |c_3|^2$. The total probability of ending up in any state is $|c_1|^2 + |c_2|^2 + |c_3|^2 = 1$. This is exactly the result we want. We now measure the spin. The probability of ending up in |II> is:

11) $\int_0^1\int_0^{2\pi}\left|c_2{}^* e^{i\theta}\sqrt{1-\beta^2}\ +\ c_3{}^*\,\beta\right|^2\ \beta\,\left(1-\beta^2\right)2/\pi\, d\theta\, d\beta = \frac{2}{3}\ |\,c_2\,|^2 + \frac{1}{3}\,|\,c_3\,|^2$

The probability of ending up in |III> is:

12) $\int_0^1\int_0^{2\pi}\left|c_2{}^* e^{i\theta}\sqrt{1-\beta^2}\ +\ c_3{}^*\,\beta\right|^2\ \beta\,\left(\beta^2\right)2/\pi\, d\theta\, d\beta = \frac{1}{3}\ |\,c_2\,|^2 + \frac{2}{3}\,|\,c_3\,|^2.$

The act of measuring the energy has cost us some information regarding the spin. We might have started with a state having $c_2 = 0$ and find the spin to be - after measuring $E_2$. This result differs from the predictions of Lüders. (The above results are relevant only if $c_1$ does not equal zero exactly and we will return to this point momentarily.) Another difference is the fact that, for us, the probabilities depend on the order of measurement — we would obtain a different result if we measured the spin first and, later, the energy. This is not the case for Lüders. We could imagine making our spin and energy measurements simultaneously. We would then recover the result of Lüders. One agreeable consequence of our method is that we are not obliged to use any particular basis for our degenerate Hilbert subspace. We have chosen the most convenient one but we could, for instance, call |II> (|A> + |B>)/$\sqrt{2}$ and |III> (|A> - |B>)/$\sqrt{2}$ or whatever we want. Our results will not change.

We could make our situation more complicated by allowing for a new state, |IV> = |Energy = $E_2$, spin = 0>. Our pure state can now be written:

13) $|\Psi_0\rangle = c_1\ |\mathrm{I}\rangle + c_2\ |\mathrm{II}\rangle + c_3\ |\mathrm{III}\rangle + c_4\ |\mathrm{IV}\rangle.$

Our most general $E_2$ eigenstate is now:

14) $\psi_a(\beta,\alpha,\theta,\delta) = e^{i\theta}\,\alpha\ |\mathrm{II}\rangle + \beta\ |\mathrm{III}\rangle +\ e^{i\delta}\sqrt{1-\beta^2-\alpha^2}\ |\mathrm{IV}\rangle$ where $\beta$ and $\alpha$ are real and between 0 and 1. $0 \le \theta, \delta \le 2\pi$.

Going through the same algebra (here we use $\mathcal{M}\,(\alpha,\beta,\theta,\delta) = \frac{6}{\pi^2}\,\alpha\,\beta$) gives us:

Probability of ending up in |II> = $\frac{1}{4}(\,2\ |\,c_2\,|^2 +\ |\,c_3\,|^2 +\ |\,c_4\,|^2)$

Probability of ending up in |III> = $\frac{1}{4}(\,|\,c_2\,|^2 + 2\ |\,c_3\,|^2 +\ |\,c_4\,|^2)$

Probability of ending up in |IV> = $\frac{1}{4}(\,|\,c_2\,|^2 +\ |\,c_3\,|^2 + 2\ |\,c_4\,|^2)$.

(If the degenerate subspace has dimension D our measure is given by $\frac{D!}{\pi^{D-1}}\ \alpha_1 ... \alpha_{D-1}$.)

Let us first measure the energy and, subsequently, the spin. We can examine the resulting change in the von Neumann entropy, - k Tr [$\rho$ ln $\rho$]. For both us and Lüders it is, of course, positive. But, according to our update rule, the entropy increase is always greater than or equal to that predicted by Lüders. We might look at this as a good thing. Physical processes generally like to increase entropy to the greatest extent possible. On the other hand, our update rule requires that "more work" be done on the initial state.

It must have struck the reader that there seems to be something strange about what we have suggested.

Consider the state given in eq. 5). If $c_1 \equiv 0$ we are already in an energy eigenstate. If the energy is measured it will certainly be found to be $E_2$. The state is therefore admissible and, according to our postulate 1), $\mathfrak{S}$ will leave it unaffected. We recover exactly the result predicted by Lüders. But, if $c_1$ differs from zero in even the smallest way, we obtain the dramatically different result described above. This discontinuous change must seem very peculiar indeed. We could say that $\mathfrak{S}$ only projects the state if $|c_1|^2$ is above some critical value. This seems arbitrary and contrived, however. Another way of achieving a desirable result would be to say that, no matter how hard we try, we just cannot, as a practical matter, ever *really* produce a state having $c_1 \equiv 0$ — there will always be a contribution, however small, from $c_1$. This is probably the best way of looking at it.

There might, actually, be a way of testing this idea. Suppose we can construct a system consisting of two massive spin-1/2 particles, a and b. We are interested in their spins. Particle a will be measured at point A and b at B. Our most general initial state can be written as eq. 13) with $|I> = |+_a, +_b>$, $|II> = |+_a, -_b>$, $|III> = |-_a, +_b>$, and $|IV> = |-_a, -_b>$. For simplicity, assume that $c_1 = c_4 = 0$. We now measure our entangled state. If A and B are spacelike separated we can always choose to work in a Lorentz frame where they are simultaneous. In this case only $|I>$, $|II>$, $|III>$, and $|IV>$ are admissible — both spins become known to the observers at the same time. Our system will be projected into either $|II>$ or $|III>$ with probabilities $|c_2|^2$ and $|c_3|^2$, respectively. We have, essentially, just re-performed something like the Aspect experiment (15) and we would get exactly his results. We think these results will always hold true provided that A and B are spacelike separated since we cannot imagine them depending on our arbitrary choice of a coordinate system. But suppose they are timelike separated with B lying inside the future light cone of A. (Aspect could not have looked at this case since his photons always travel at *c*. He might, however, have measured his first photon then, using a mirror, reflected the second photon back to the polarimeter to be measured later.) Suppose a is measured at A and found to be +. According to Lüders the state is now $|II>$ and we are certain to find b - when it is later measured. For us the first measurement can project the state into any linear combination of $|I>$ and $|II>$ (as described above). Thus, when b is measured at B, there will only be a 2/3 chance of finding it -. It could be +.

## Brains as Detectors.

Most physicists would probably rather believe that it is a property of their instrumental detectors that refuses to exist in some kinds of superpositions. Perhaps it is their size or complexity? But rocks can be fairly large and complex too. It is hard to see how they collapse wave functions. Also, how to explain the fact that, when wave functions do collapse, they always seem collapse in such a manner that consciousness remains in a definite state and does not wander into an absurd one? This seems peculiar. And, since quantum measurements, as we understand them, appear to require the intervention of experimental physicists, would the universe have to wait for billions of years for animals to develop quantum mechanics, bubble chambers, and photomultiplier tubes before wave functions could collapse at all? Maybe, but this too seems somewhat peculiar.

We explore a different possibility. We do believe in psychophysiological parallelism – we think there are physical brain processes (embodied in $|\Psi(t)>$) that correlate with our qualia-states. Suppose these processes are exquisitely sensitive to some, irreducibly quantum mechanical, effects of the brain's structure or function. Essentially, the brain is a sort of quantum computer (albeit completely different from any that have been constructed in laboratories). Various candidates (e.g. microtubules (16), catecholaminergic neurotransmission (17), Integrated Information Theory (5), and the possible quantum mechanical behavior of synapses (18)) have been proposed. But we note that many such processes are likely to be sensitive to quantum decoherence. The brain is a "warm, wet, noisy environment." Quantum decoherence represents a serious problem for the designers of quantum computers and for, at least, the microtubule hypothesis as well (19).

Let us consider a very simple universe – one where $|\Psi(t)>$ describes only a single, conscious, brain floating in a nutrient bath and experiencing qualia. $|\Psi(t)>$ begins in an admissible state. But it might very well decohere (through unitary evolution) into an inadmissible one. At the moment this tries to happen $\mathfrak{S}$ projects it back into an admissible state as described above. It may then evolve smoothly until it, again, begins to decohere. $\mathfrak{S}$ will then project it once more. Whether this decoherence sets in every femtosecond, every day, or every year we cannot say. But, if this view is correct, $\mathfrak{S}$ exists not so much to clean up the results of Stern-Gerlach experiments and such as it does to *enable* conscious brains to function as workable quantum computers. $\mathfrak{S}$ is not simply a consequence of consciousness but, also, an *absolute precondition for consciousness*. A universe without $\mathfrak{S}$ might be able to accommodate very simple animals, assuming their unconscious brains functioned at a completely classical level. But it could never generate interesting, conscious, animals. This observation does not depend on the exact manner in which $\mathfrak{S}$ accomplishes its task. It could, for instance, function in the manner Chalmers and McQueen suggest and be just as effective in protecting $|\Psi(t)>$ from decoherence. (It is unfortunate that we have, apparently, not been able to manufacture a conscious quantum computer. If we could $\mathfrak{S}$ would, presumably, take care of its decoherence problems!)

## Why Consciousness?

As far as we know, the universe began with a brief moment of inflation during which $|\Psi(t)>$, living in the Fock space of the Standard Model, became populated with the enormous number of elementary particles we now see. This could, presumably, have happened in any number of ways depending on the nature of the inflaton field and the manner of its transition from its false vacuum to its true one. So our universe could have begun in a great many initial states. We will designate these $|\Psi_i(\epsilon)>$ (where $\epsilon$ is the very small time over which the inflation occurred). We assume they are all admissible null states. We expect that many of these initial states would be completely inconsistent with the emergence of consciousness – they might give rise to universes that were too hot or cold or dense or diaphanous or homogeneous or inhomogeneous to ever permit the formation of stars and planets. No conscious life could ever evolve. But some (one, at least) must have been such as to allow for this. We know that because we are, obviously, here. We will call the first set 'sterile' and the other 'fertile.'

Now sterile and fertile initial conditions would evolve quite differently. The former would simply evolve from one admissible null state to another to another forever. $\mathfrak{S}$ would do nothing. Each would give rise to a single world-history $|\Psi_i(t)>$. In a fertile universe conscious animals would, eventually, appear and start making quantum measurements. Every time this happened the world-history would diverge into a (possibly very large) number of new branches. In time, each of these will branch still further. As $t \to \infty$ the probabilities of the various results of the measurements must, of course, converge to the values predicted by Quantum Mechanics. Suppose our branches come about through repeatedly performing a quantum measurement that has a 60% chance of giving a 'green' result and a 40% chance of a 'red' one. Any acceptable branch will, in the limit, be found to contain 1.5 times as many 'green' nodes as 'red' ones. And, if we look at sections of a branch, we will usually find more sequences like ...GGGGG... than ...RRRRR... . An infinite sequence of all Gs would not constitute a valid world-history since it would not satisfy the predictions of Quantum Mechanics.

We will introduce a new pictorial metaphor. The world-history deriving from a sterile initial condition would look like a single, infinitely long, blade of grass – it would never branch at all. One deriving from a fertile initial condition would look like a dense, complicated, *bush*. It would consist of infinitely many world-histories each being perfectly consistent with the laws of physics. And not all fertile $|\Psi_i(\epsilon)>$s are created equal.

Some may be very conducive to the evolution of consciousness – in these many quantum measurements would take place. Others might give rise to only a few conscious animals and these might die out quickly. We would end up with a bush in either case. In the second case the bush would be small and sparse. In the first, it would be a great deal "bushier."

Now we may think that there is one, and only one, real world-history. But we also suspect that sterile initial conditions must be far more common than fertile ones. (It is easier to design a watch that does not work than it is to design one that does.) But we are here and conscious.

We could make an appeal to Anthropic Principles, but we will do something completely different. For lack of any reason to think otherwise, let us assume that the real universe is picked at random out of all the conceivable valid world-histories. This is the simplest way things could be. After all, any world-history will satisfy the laws of physics equally well. It is then overwhelmingly likely that the one picked will belong to a bush and not be a blade of grass. In fact, it is overwhelmingly likely to belong to the bushiest part of the bushiest bush possible given the laws of physics. If what we have said so far is correct, our universe is overwhelmingly likely to have evolved from the most fertile initial condition possible. Thus we do not have to regard the existence of consciousness in Nature as a happy accident or the result of Divine Intervention or some Anthropic Principle. Rather, it is just an ineluctable, statistical, consequence of the way things work. And this conclusion is, also, fairly robust. Perhaps we are wrong about how $\mathfrak{S}$ works. Maybe it works like Chalmers and McQueen say, or in some other way. But, if it branches world-histories, we will still end up with bushes and blades of grass. Or, perhaps, Everett is right. In this case all possibly conscious world-histories *would* be conscious. "We" would still be vastly more likely to find "ourselves" in a bushy branch in a bushy bush rather than elsewhere.

## Acknowledgement.

The author is grateful to Professors David J. Chalmers, Lawrence S. Schulman, Saverio Pascazio, and Anthony Sudbery for useful and interesting discussions regarding this work.

## References and Footnotes.